\documentclass[prl,twocolumn,showpacs]{revtex4}

\usepackage{graphicx}
\begin{document}

\title{
   Quasiexcitons in Incompressible Quantum Liquids}

\author{
   Arkadiusz W\'ojs,$^{1,2}$ 
   Anna G{\l}adysiewicz,$^2$ and 
   John J. Quinn$^1$}

\affiliation{
   $^1$University of Tennessee, Knoxville, Tennessee 37996, USA\\
   $^2$Wroclaw University of Technology, 50-370 Wroclaw, Poland}

\begin{abstract}
Photoluminescence (PL) has been used to study two-dimensional 
incompressible electron liquids in high magnetic fields for nearly 
two decades.
However, some of the observed anomalies coincident with the fractional 
quantum Hall effect are still unexplained.
We show that emission in these systems occurs from fractionally 
charged ``quasiexciton'' states formed from trions correlated 
with the surrounding electrons.
Their binding and recombination depend on the state of both the 
electron liquid and the involved trion, predicting discontinuities 
in PL and sensitivity to sample parameters.
\end{abstract}
\maketitle

The density of states of a two-dimensional electron gas (2DEG) in 
a high magnetic field $B$ consists of discrete Landau levels (LL's).
When all electrons fall into the lowest LL, it is their mutual 
interaction that solely determines the ground state (GS) and 
low-energy excitations.
Reminiscent of atomic physics except for the macroscopic LL 
degeneracy, this makes the 2DEG at high $B$ an ideal laboratory 
of many-body physics in extended systems.

The incompressible quantum liquids (IQL's) \cite{Laughlin83} were 
originally discovered in transport experiments \cite{Tsui82} over 
two decades ago, but it took 15 years to demonstrate their hallmark 
fractionally charged quasiparticles (QP's) by shot-noise studies 
\cite{Saminadayar97}.
Photoluminescence (PL) was also used to probe IQL's, revealing 
anomalies at the LL fillings $\nu$ coincident with fractional 
quantum Hall effect \cite{Heiman88,Goldberg90,Turberfield90,Plentz96,%
Byszewski04} (usually doublets at $\nu\approx{1\over3}$ or ${2\over3}$,
but other features, too \cite{Yusa01,Schuller03}).
Other optical experiments include PL with acceptor-bound holes 
\cite{Kukushkin94} and Raman scattering \cite{Pinczuk93}.

The connection of PL anomalies with the microscopic properties 
of IQL's has been studied theoretically for over a decade.
Depending on the strength and resolution of the Coulomb potential 
of photo-injected holes (controlled by $h$--2DEG separation 
\cite{Plentz96}), the observed doublets were attributed either 
to the ``bare exciton'' and ``magnetoroton assisted'' emission 
\cite{Apalkov92,Apalkov95} (efficient due to ``gigantic suppression 
of the exciton dispersion by an IQL'' \cite{Apalkov95}), or to 
recombination of different ``anyon excitons'' \cite{Rashba93,Chen93} 
consisting of several fractional IQL QP's bound to a hole.
However, understanding of all reported anomalies is not yet complete and, 
e.g., discontinuities reported in Refs.~\cite{Goldberg90,Byszewski04} 
remain, to our best knowledge, unexplained.

To appreciate the complexity of the problem, one must recall that: 
(i) Even an unperturbed IQL has complicated dynamics whose understanding 
involves concepts of Laughlin correlations and fractionally charged QP's 
\cite{Laughlin83}, anyon statistics \cite{Halperin84}, Haldane hierarchy 
\cite{Haldane83}, or composite fermions (CF's) \cite{Jain89}.
(ii) Emergence of ``multiplicative states'' in $e$--$h$ fluids with 
``hidden symmetry'' (HS) \cite{Lerner81} greatly simplifies their optical 
response.
(iii) Breaking of HS in real systems (due to finite layer widths $w$,
charge separation, LL and valence band mixing, or disorder) 
restores the possibility of IQL-related anomalies in PL.

The HS-related effects in real quantum wells are well known in the ``dilute'' 
regime ($\nu\ll1$), in which PL is determined by recombination of excitons 
(${\rm X}=e+h$) and trions (${\rm X}^-=2e+h$) \cite{Yusa01,Kheng93}.
HS precludes radiative complexes larger than X, allowing for only one trion 
-- ``dark triplet'' X$^-_{\rm t}$ \cite{TD}.
It is only due to the LL mixing that a ``bright singlet'' X$^-_{\rm s}$ 
occurs as well \cite{Stebe89,SB}.

In the ``liquid'' regime, few-body excitonic effects compete with 
many-body IQL dynamics, adding to each one's own complexity almost 
to guarantee fascinating physics.
Different photo-excitations weakly coupled to the remaining IQL were 
proposed. 
In the anyon exciton model \cite{Rashba93,Chen93} applicable for 
structures with strong charge separation (heterojunctions or wide 
asymmetric wells), the holes repel quasiholes (QH's) and attract 
quasielectrons (QE's) of the IQL.
The ``dressed exciton'' concept \cite{Apalkov92,Apalkov95} introduced 
for narrower wells, involves the X's coupled to magnetorotons of the 
IQL.
In another approach \cite{Chen93,X-CF} the X$^-$'s correlate with 
the surrounding electrons.

In this Letter we develop the idea of trions immersed in a Laughlin IQL 
and {\em predict discontinuity of the PL spectrum} at $\nu={1\over3}$ 
\cite{Goldberg90,Byszewski04}.
We show that trions remain stable in realistic doped wells, but acquire 
effective charge $\mathcal{Q}$ of up to one ``Laughlin quantum'' 
$\varepsilon={1\over3}e$ due to partial screening by the IQL.
In analogy with X and X$^\pm$, we find {\em neutral and charged 
``quasiexcitons''} (QX's): $\mathcal{X}$ and $\mathcal{X}^{\pm1/3}$.
They consist of a trion which is Laughlin-correlated with the IQL 
and binds 0, 1, or 2 QH's.
The $\mathcal{X}^{\pm1/3}$ binding energies are directly observable 
in PL, and their order-of-magnitude reduction from the X$^\pm$ is 
an {\em indication of the fractional charge} of their constituents.

For spin-polarized systems, we elucidate the earlier theory 
\cite{Apalkov92,Apalkov95} by identifying the ``dressed exciton'' with 
$\mathcal{X}$, its suppressed dispersion with the $\mathcal{X}^{-1/3}$--QH 
pseudopotential, and the ``magnetoroton-assisted emission'' with the 
$\mathcal{X}^{-1/3}$ recombination. 
The PL discontinuity proposed here due to charged QX's is a different
effect, requiring no thermal activation and no charge separation.

In unpolarized systems, we find a spin-flip $\mathcal{X}$ whose steep 
dispersion prevents charging and removes the PL discontinuity.
Competition between X$^-_{\rm s}$ and X$^-_{\rm t}$ in realistic 
wells {\em predicts dependence of PL anomalies on $w$}.

We use exact numerical diagonalization for $N\le10$ electrons and one 
hole on a sphere \cite{Haldane83} (with radius $R$, monopole strength 
$2Q=4\pi R^2 Be/hc$, magnetic length $\lambda=R/\sqrt{Q}$).
Interaction matrix elements were integrated taking subband wavefunctions 
$\phi_{e,h}(z)$ calculated self-consistently \cite{Tan90} for $w=10$ 
and 20~nm GaAs wells, doped on one side to $n=2\cdot10^{11}$~cm$^{-2}$ 
(yielding $\nu={1\over3}$ at $B=25$~T, the values used further on).
Finite size errors were minimized by extrapolation to 
$\lambda/R\rightarrow0$.

We begin with calculation of X$^-$ Coulomb binding energies $\Delta$ 
using $\phi_{e,h}$, i.e., in mean normal electric field but ignoring 
in-plane X$^-$-IQL coupling.
We included five LL's and two $\phi$-subbands for both $e$ and (heavy) 
$h$, but neglected valence subband mixing.
For $w=10$~nm, $\Delta_{\rm s}=2.3$~meV and $\Delta_{\rm t}=1.5$~meV, 
in qualitative agreement with Refs.~\cite{Stebe89,SB}, also predicting 
the X$^-_{\rm s}$ ground state (GS).
For $w=20$~nm, neither symmetric-well nor lowest-subband approximation
works well (e.g., the latter exaggerates charge separation in X/X$^-$ 
and predicts breakup of X$^-_{\rm s}$).
Our best estimates, $\Delta_{\rm s}=1.5$~meV and $\Delta_{\rm t}=
1.2$~meV, are rather sensitive to the parameters, making prediction 
of the X$^-$ GS in real samples difficult and somewhat pointless.
However, we expect that the X$^-_{\rm t}$'s, additionally favored by 
the Zeeman energy, could at least coexist with the X$^-_{\rm s}$'s at 
finite temperatures.

Let us immerse a trion (X$^-_{\rm s}$ or X$^-_{\rm t}$, depending on 
$w$, $n$, and $B$) in an IQL.
Effective $e$--X$^-$ pseudopotentials are similar \cite{SB} to the 
$e$--$e$ one \cite{Haldane87}.
In the lowest LL, this causes similar $e$--$e$ and $e$--X$^-$ 
correlations, described in a generalized two-component \cite{X-CF} 
CF picture \cite{Jain89} by attachment of $2p$ flux quanta to each 
$e$ and X$^-$.
At Laughlin/Jain fillings $\nu_{\rm IQL}=s/(2ps+1)$, electrons 
converted to CF$_e$'s fill the lowest $s$ LL's in effective magnetic 
field $B^*=B-2pn(hc/e)=B/(2ps+1)$.
At $\nu\ne\nu_{\rm IQL}$, QE's in the $(s+1)^{\rm st}$ or QH's in 
the $s^{\rm th}$ CF$_e$-LL occur, carrying effective charge 
$\varepsilon=\pm e/(2ps+1)$.
We find that, similarly, an X$^-$ is converted to a CF$_{\rm X^-}$ 
with charge $\mathcal{Q}=-\varepsilon$.

A trion coupled to an IQL and carrying reduced charge is a 
many-body excitation.
To distinguish it from an isolated $2e+h$ state, we call it 
a charged {\em quasiexciton} (QX) and denote by $\mathcal{X}^-
\equiv\mathcal{X}^{-\varepsilon}$.
Being negatively charged, an $\mathcal{X}^-$ interacts with IQL QP's. 
At $\nu<\nu_{\rm IQL}$, the $\mathcal{X}^-$ binds a QH to
become a neutral $\mathcal{X}^-{\rm QH}=\mathcal{X}$, with 
a binding energy called $\Delta^0$.
Depending on sample parameters and spin of the trion, $\mathcal{X}$ 
may bind an additional QH to form a positively charged 
$\mathcal{X}^-{\rm QH}_2=\mathcal{X}^+$, with binding energy $\Delta^+$.
At $\nu>\nu_{\rm IQL}$, the $\mathcal{X}^+$ attracts and annihilates 
a QE: $\mathcal{X}^++{\rm QE}\rightarrow\mathcal{X}$; this process 
releases energy $\Delta_{\rm IQL}-\Delta^+$ (where $\Delta_{\rm IQL}=
\mathcal{E}_{\rm QE}+\mathcal{E}_{\rm QH}$ is the IQL gap).
The $\mathcal{X}$ may annihilate another QE: $\mathcal{X}+{\rm QE}
\rightarrow\mathcal{X}^-$, with energy gain
\begin{equation}
   \Delta^-=\Delta_{\rm IQL}-\Delta^0
\label{eq1}
\end{equation}
that can be interpreted as $\mathcal{X}^-$ binding energy.

The $\mathcal{X}$ and $\mathcal{X}^\pm$ are different states in which 
a hole can exist in an IQL.
If $\Delta^\pm>0$, then depending on $\nu$, either $\mathcal{X}^-$ or 
$\mathcal{X}^+$ is the most strongly bound state.
If $\Delta^-\ne\Delta^+$, the PL spectrum will be discontinuous at 
$\nu_{\rm IQL}$.
For long-lived $\mathcal{X}^\pm$ (made of a dark X$^-_{\rm t}$), 
recombination of the $\mathcal{X}$ is also possible, especially 
at $\nu\approx\nu_{\rm IQL}$ (within a Hall plateau), when QP 
localization impedes $\mathcal{X}^\pm$ formation.

The QX's resemble normal excitons in $n$- or $p$-type systems, 
except that the concentration of their constituent QP's can be 
varied (in the same sample) by a magnetic field.
Also, their kinetics ($\mathcal{X}\leftrightarrow\mathcal{X}^\pm$) 
is more complicated because of the involved QE--QH annihilation.

\begin{figure}
\resizebox{3.4in}{3.33in}{\includegraphics{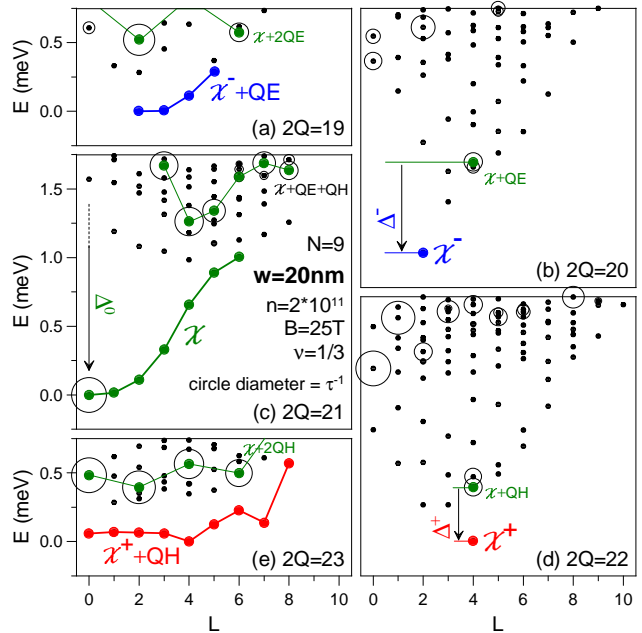}}
\caption{(color online)
   Excitation energy spectra (energy $E$ as a function of total 
   angular momentum $L$) of $9e+h$ systems on a sphere, with up 
   to two QE's or QH's in Laughlin $\nu={1\over3}$ IQL.
   Oscillator strengths $\tau^{-1}$ are indicated by open circles.}
\label{fig1}
\end{figure}

We have tested the QX idea numerically for Laughlin $\nu={1\over3}$ IQL.
First, we calculated spin-polarized $Ne+h$ energy spectra for $w=20$~nm,
in search of the QX$_{\rm t}$'s.
The X$^-_{\rm t}$ has 94\% squared projection onto the lowest LL, so
we ignored LL mixing in the $Ne+h$ calculation (direct tests confirmed 
that it is negligible).
The low-lying states in Fig.~\ref{fig1} are understood using the CF 
picture \cite{Jain89,X-CF} and addition rules for angular momentum.
On a sphere, the CF transformation introduces an effective monopole 
strength $2Q^*=2Q-2(K-1)$, where $K=N-1$ is the total number of free 
electrons and X$^-$'s.
The angular momenta of constituent QP's are $l_{\rm QH}=Q^*$, 
$l_{\rm QE}=Q^*+1$, and $l_{\mathcal{X}^-}=Q^*-1$.
The $\mathcal{X}^-$ is a dark GS in (b) at $L=l_{\mathcal{X}^-}=2$, 
and $\mathcal{X}^+$ is found in (d) at $L=l_{\mathcal{X}^+}=
|(2l_{\rm QH}-1)-l_{\mathcal{X}^-}|=4$.
Bands of $\mathcal{X}^-$--QE and $\mathcal{X}^+$--QH pairs are 
marked in (a) and (e).
In (c) the radiative $L=0$ GS is a multiplicative state, opening 
a $\mathcal{X}=\mathcal{X}^-$--QH band \cite{Chen93}, earlier called 
a ``dressed exciton'' and identified \cite{Apalkov92,Apalkov95} as 
responsible for the doublet structure in PL.

\begin{figure}
\resizebox{3.4in}{1.25in}{\includegraphics{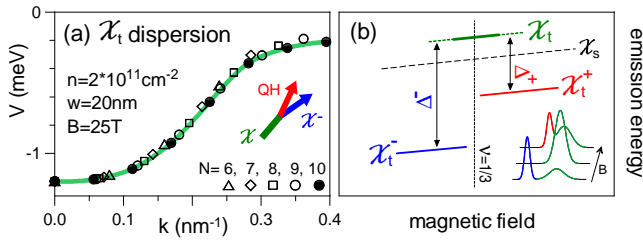}}
\caption{(color online)
   (a) Dispersion of neutral quasiexciton $\mathcal{X}_{\rm t}$
   in Laughlin $\nu={1\over3}$ IQL;
   $\mathcal{X}_{\rm t}$ splits into $\mathcal{X}^-_t$ and QH 
   at $k>0$.
   (b) Schematic of PL discontinuity due to 
   $\mathcal{X}_{\rm t}^\pm$ emission.}
\label{fig2}
\end{figure}

The continuous $\mathcal{X}$ dispersion shown in Fig.~\ref{fig2}(a)
results \cite{Apalkov92,Apalkov95} from the in-plane dipole moment 
being proportional to the wavevector $k=l/R$.
Here we find why it is suppressed (compared to X): because of the 
reduced charge of the $\mathcal{X}$'s constituents, $\mathcal{X}^-_t$ 
and QH.
In fact, the $\mathcal{X}$ and X dispersions become similar when 
energy and length scales are rescaled in account of the $e\rightarrow
\varepsilon$ charge reduction.
We also explain the emission from $\mathcal{X}$ at $k\lambda\sim1.5$, 
proposed \cite{Apalkov92,Apalkov95} for the lower peak in PL, as 
the $\mathcal{X}^-\rightarrow{\rm QE}$ recombination assisted by 
QH scattering. 
However, note that a small $dV/dk$ and a large $\tau^{-1}$ at $k\lambda
\sim1.5$ needed for this emission requires $w>20$~nm.

By identifying the multiplicative states containing an $\mathcal{X}$
with $k=0$, one can estimate $\Delta^{\pm/0}$ as marked in 
Fig.~\ref{fig1}(b)--(d).
More accurate values were obtained by comparing spectra in which
$\mathcal{X}^\pm$, $\mathcal{X}$, or QP is alone in the IQL, 
followed by extrapolation to $N\rightarrow\infty$:
$\mathcal{E}_{\rm QH}=0.73$~meV, 
$\mathcal{E}_{\rm QE}=1.05$~meV,
$\Delta^0=1.20$~meV,
$\Delta^-=0.52$~meV, and
$\Delta^+=0.27$~meV.
Depending on $\mathcal{X}^0/\mathcal{X}^\pm$ kinetics, either 
$\Delta^+\ne\Delta^-$ or $\Delta^0\ne\Delta^\pm$ asymmetry will make 
PL energy jump at $\nu={1\over3}$, as sketched in Fig.~\ref{fig2}(b).
Similar behavior has been observed \cite{Goldberg90,Byszewski04}.

\begin{figure}
\resizebox{3.4in}{1.25in}{\includegraphics{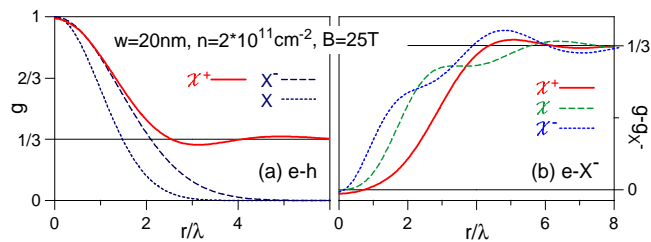}}
\caption{(color online)
   (a) $e$--$h$ PDF's $g(r)$ of quasiexciton $\mathcal{X}^+_{\rm t}$ 
   and isolated X$^-_{\rm t}$ and X, normalized to measure local filling 
   factor.
   (b) $e$--X$^-$ PDF's for different QX's;
   curve for $\mathcal{X}^+$ resembles $e$--$e$ PDF of Laughlin 
   liquid; shoulders for $\mathcal{X}$ and $\mathcal{X}^+$ reflect 
   additional charge quanta pushed onto the hole.}
\label{fig3}
\end{figure}

The $\mathcal{X}^\pm$ discontinuity is different from that due to 
anyon excitons \cite{Rashba93,Chen93} anticipated in much wider wells.
The two can be distinguished by different magnitude 
(${\sim}\Delta_{\rm IQL}$ vs.\ $\Delta^\pm$) and opposite direction 
of the jump of emission energy when passing through $\nu={1\over3}$.
In the present case, the small ratio of $\mathcal{X}^\pm$ and X$^\pm$ 
binding energies is the {\em signature of the fractional charge} of 
the IQL excitations -- directly observable as splittings in PL.

The QX's are defined through a sequence of gedanken processes:
(i) trion binding: $2e+h\rightarrow{\rm X}^-$,
(ii) Laughlin correlation: X$^-\rightarrow\mathcal{X}^-$, 
(iii) QH capture: $\mathcal{X}^-\rightarrow\mathcal{X}/\mathcal{X}^+$.
Hence, $\mathcal{X}$ and $\mathcal{X}^\pm$ are the same X$^-$, 
but differently separated from the surrounding electrons.
This is evident in $e$--$h$ pair-distribution functions (PDF's)
in Fig.~\ref{fig3}.
Integration of $g(r)$ confirms fractional electron charge 
$-{4\over3}e$, $-e$, and $-{2\over3}e$ bound to the hole in the 
$\mathcal{X}^-$, $\mathcal{X}$, and $\mathcal{X}^+$.

Let us now turn to spin-unpolarized spectra, in search of QX's formed 
from the X$^-_{\rm s}$.
Its binding depends on LL mixing, so we used the following approximation.
In the calculation of Coulomb matrix elements, the highest Haldane 
$e$--$e$ pseudopotential, $V_0$, was reduced by 10\%.
This only affects interactions in the trion, and induces an (85\% accurate) 
X$^-_{\rm s}$ GS in the lowest LL.
It has correct $g_{eh}$ and $g_{ee}$ PDF's, which determine coupling 
to the 2DEG.

\begin{figure}
\resizebox{3.4in}{3.33in}{\includegraphics{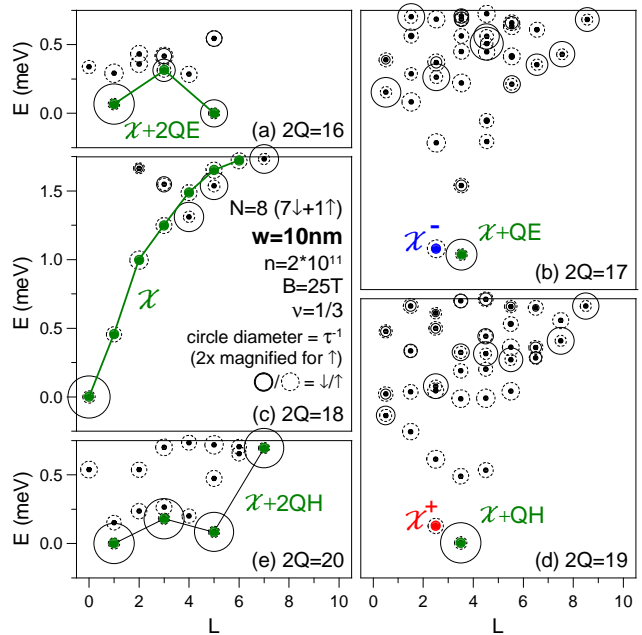}}
\caption{(color online)
   Excitation spectra similar to Fig.~\ref{fig1}, but for the 
   $8e+h$ systems with one reversed-spin electron and $w=10$~nm.
   $\tau^{-1}$ is separate for spin-$\downarrow$ and $\uparrow$ 
   recombination.}
\label{fig4}
\end{figure}

From the analysis of $\tau^{-1}$ and $L$ we found $\mathcal{X}_{\rm s}$ 
and $\mathcal{X}_{\rm s}^\pm$ in the spectra in Fig.~\ref{fig4}.
In contrast with Fig.~\ref{fig1}, charged QX$_{\rm s}$'s are the 
excited states at $2Q=17$ and 19.
Also at $2Q=16$ and 20, the multiplicative states with an 
$\mathcal{X}_{\rm s}$ and two QP's lie below the 
$\mathcal{X}^-_{\rm s}$--QE and $\mathcal{X}^+_{\rm s}$--QH pairs.

This opposite behavior results from the X$^-_{\rm s}$ having 
different charge distribution than the X$^-_{\rm t}$.
It has little effect on its Laughlin correlation with the electrons, 
but affects its interaction with the QP's.
Indeed, the $\mathcal{X}_{\rm s}$ dispersion in Fig.~\ref{fig4}(c) 
indicates stronger $\mathcal{X}^-_{\rm s}$--QH attraction.

We compare $\Delta^0\sim2$~meV with $\Delta_{\rm IQL}=2.02$~meV 
using Eq.~(\ref{eq1}) to find that $\Delta^-$ is very small or even 
negative, as in Fig.~\ref{fig4}(b).
Hence, even in the absence of free QH's, the $\mathcal{X}^-_{\rm s}$ 
is unstable toward creation of a QE--QH pair.
Similarly, negative $\Delta^+$ in Fig.~\ref{fig4}(d) implies 
instability of the $\mathcal{X}^+_{\rm s}$.
As a result, the neutral $\mathcal{X}_{\rm s}$ is the most strongly 
bound state regardless of the presence of QE's or QH's.

This may add a continuous peak for the $w=20$~nm well 
[see Fig.~\ref{fig2}(b)], but precludes PL discontinuity in narrow 
wells with a strong X$^-_{\rm s}$ GS.
The $\mathcal{X}_{\rm s}$ peak splits into a $\sigma_\pm$ doublet 
due to spin-$\downarrow$ and $\uparrow$ recombination involving 
either QE's or ``reversed-spin'' QE$_{\rm R}$'s \cite{Rezayi87}, 
but temperature-activated emission at $k>0$ is not expected.

The QX idea can be extended to other IQL's (e.g., $\nu={2\over3}$ or 
${2\over5}$).
However, different behavior of QX$_{\rm t}$'s and QX$_{\rm s}$'s at 
$\nu={1\over3}$ is an example that PL discontinuity is not guaranteed.
Via Eq.~(\ref{eq1}), it is governed by sample- and $\nu$-dependent 
$\Delta_{\rm IQL}$ and $\Delta^0$ which must be recalculated.

In summary, we have studied anomalies in PL of the IQL's in the 
regime of small charge separation.
The emission spectrum is due to recombination of QX's formed from 
trions immersed in a 2DEG with Laughlin correlations.
In spin-polarized systems, the neutral QX is equivalent to a nearly 
decoupled exciton at $k=0$, and its suppressed dispersion results 
from reduced charge of the constituents.
The positive and negative spin-polarized QX's have fractional charge 
of one IQL QP.
A spin-flip QX formed from a singlet trion was also found, with 
a steeper dispersion that prevents its charging by the IQL QP's.
Featureless PL of the IQL in narrow (10~nm) wells and anomalies 
predicted for wider (20~nm) wells agree qualitatively with experiments.

We thank M.~Potemski and P.~Hawrylak for helpful discussions and 
sharing their results prior to publication.
Work supported by grants DE-FG 02-97ER45657 of US DOE and 
2P03B02424 of Polish MENiS.

\end{document}